# The Diamine Cation Is Not a Chemical Example Where Density Functional Theory Fails

Zulfikhar A. Ali[1], Fredy W. Aquino[2], and Bryan M. Wong[1,2]*

[1]Department of Physics & Astronomy, University of California-Riverside, Riverside, California 92521, United States

[2]Department of Chemical & Environmental Engineering and Materials Science & Engineering Program, University of California-Riverside, Riverside, California 92521, United States

*Corresponding author. E-mail: bryan.wong@ucr.edu, http://www.bmwong-group.com

In a recent communication, Weber and co-workers[1] presented a surprising study on charge-localization effects in the *N,N'*-dimethylpiperazine (DMP$^+$) diamine cation to provide a stringent test of density functional theory (DFT) methods. Within their study, the authors examined various DFT methods and concluded that "all DFT functionals commonly used today, including hybrid functionals with exact exchange, fail to predict a stable charge-localized state."[1] This surprising conclusion is based on the authors' use of a self-interaction correction (namely, complex-valued Perdew-Zunger Self-Interaction Correction (PZ-SIC))[2,3] to DFT, which appears to give excellent agreement with experiment and other wavefunction-based benchmarks. Since the publication of this recent communication, the same DMP$^+$ molecule has been cited in numerous subsequent studies[4-13] as a prototypical example of the importance of self-interaction corrections for accurately calculating other chemical systems. In this correspondence, we have carried out new high-level CCSD(T) analyses on the DMP$^+$ cation to show that DFT actually performs quite well for this system (in contrast to their conclusion that all DFT functionals fail), whereas the PZ-SIC approach used by Weber et al. is the outlier that is inconsistent with the high-level CCSD(T) (coupled-cluster with single and double excitations and perturbative triples) calculations. Our new findings and analysis for this system are briefly discussed in this correspondence.

In Figure 1, we re-plot the PZ-SIC and M06-HF (Minnesota 06 with Hartree Fock exchange) potential energy curves (reproduced from Ref. 1 by Weber et al.) overlaid on top of our new MP2 (Møller-Plesset $2^{nd}$ order perturbation theory), CCSD (coupled-cluster with single and double excitations), and CCSD(T) calculations. Using the same nomenclature as Ref. 1, the charge-delocalized dimelthylpiperazine (DMP-D$^+$) structure occupies the global minimum on the potential energy curve and is characterized by a positive charge that is delocalized over the two equivalent nitrogen atoms. In contrast, the charge-localized dimelthylpiperazine (DMP-L$^+$) structure occupies a local minimum on the potential energy curve and has a positive charge that is localized on only one of the nitrogen atoms. The CCSD_CCSD(T)-SP and MP2_CCSD(T)-SP legend labels in Fig. 1 denote single-point (SP) energy calculations that were carried out with the CCSD(T) method using geometry-optimized structures obtained with CCSD and MP2, respectively. To maintain a consistent comparison with the previous study by Weber et al., the same basis sets from Ref. 1 were used throughout this work (i.e., all optimizations were carried out with the aug-cc-pVDZ (augmented correlation-consistent polarized valence double-zeta) basis set, and single-point energy CCSD(T) calculations utilized the cc-pVTZ (correlation-consistent polarized valence triple-zeta) basis). It is worth noting that Weber and co-workers did not examine any details of the transition state structure using high-level wavefunction-based calculations, which we provide for the first time in both Fig. 1 and the Supplementary Figs. 1 and 2. Upon examination of Fig. 1, we observe several clear trends. First, all three wavefunction-based approaches (CCSD, CCSD_CCSD(T)-SP, and MP2_CCSD(T)-SP) are in agreement by producing a potential energy curve with an extremely small energy barrier (< 0.01 eV), which is in stark contrast to the much larger 0.2 eV barrier obtained from the PZ-SIC approach. Most interestingly, a single-point CCSD(T) energy calculation on top of the CCSD- and MP2-optimized geometries further lowers the barrier to the point where it more closely resembles the M06-HF potential energy curve. While we take the CCSD_CCSD(T)-SP curve in Fig. 1 to be the most accurate calculation among all the methods studied, it is interesting to note that the MP2_CCSD(T)-SP curve still closely resembles both the CCSD_CCSD(T)-SP and M06-HF curves. In addition to the barrier height, the CCSD(T) single-point calculations alter the relative energy difference between the DMP-D$^+$ and DMP-L$^+$ structures such that CCSD_CCSD(T)-SP and MP2_CCSD(T)-SP curves are even closer in agreement with the M06-HF DFT calculations. Table 1 summarizes the barrier heights and relative energy differences obtained from PZ-SIC, M06-HF, and the various wavefunction-based methods examined. Taken together, both the small barrier heights and the DMP-D$^+$/DMP-L$^+$ relative energy differences obtained from the high-level CCSD(T) calculations show good agreement with the DFT methods examined in Ref. 1, and it is actually the PZ-SIC calculation that is the outlier and inconsistent with the highly-accurate CCSD(T) benchmarks.

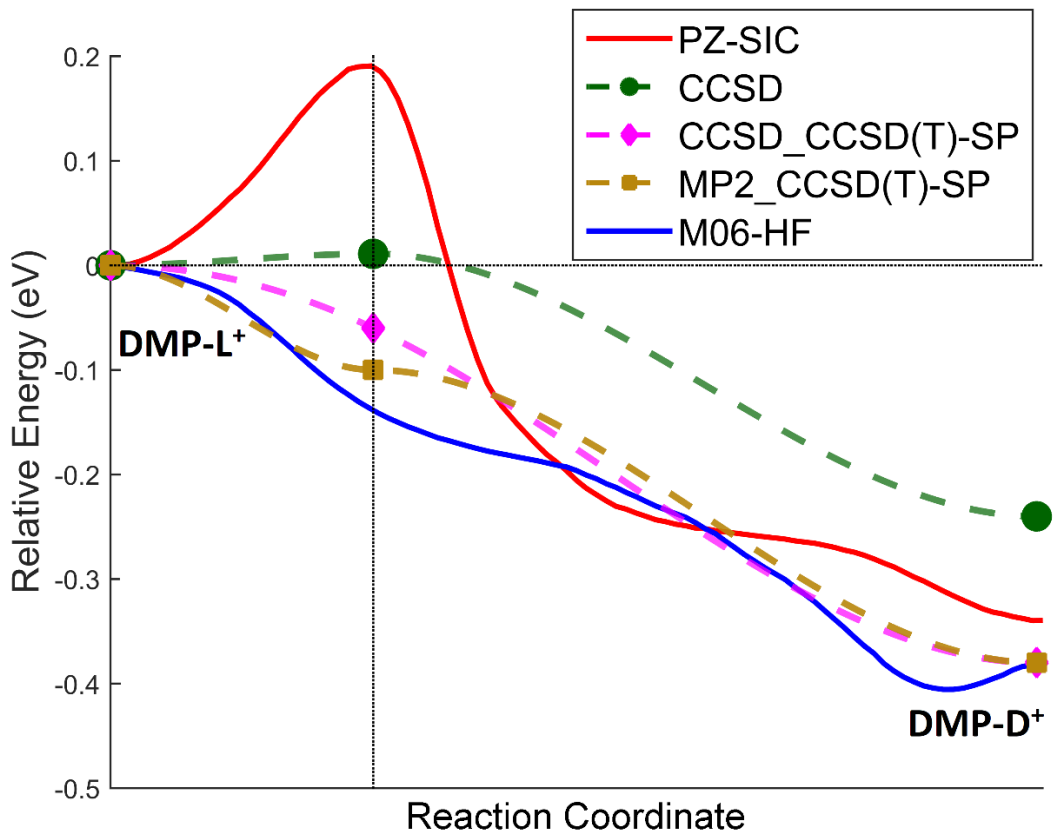

**Figure 1 | Calculated potential energy curve between the localized and delocalized state of the dimelthylpiperazine cation.** The PZ-SIC and M06-HF potential energy curves were obtained from Ref. 1, and the CCSD_CCSD(T)-SP and MP2_CCSD(T)-SP legend labels denote single-point energy calculations that were carried out with the CCSD(T) method using geometry-optimized structures obtained with CCSD and MP2, respectively. All three wavefunction-based approaches (CCSD, CCSD_CCSD(T)-SP, and MP2_CCSD(T)-SP) are in agreement by producing an extremely small energy barrier (< 0.01 eV), with the CCSD_CCSD(T)-SP and MP2_CCSD(T)-SP curves in close agreement with the M06-HF DFT calculations.

**Table 1 | Relative energies of the DMP-L$^+$ and DMP-D$^+$ states obtained from various computational methods.**

| Method | Relative Energy (eV) | |
|---|---|---|
| | Barrier Height | Energy (DMP-L$^+$) – Energy (DMP-D$^+$) |
| PZ-SIC | 0.20 | 0.34 |
| M06-HF | 0.00 | 0.38 |
| CCSD | 0.01 | 0.24 |
| MP2_CCSD(T)-SP | 0.00 | 0.38 |
| CCSD_CCSD(T)-SP | 0.00 | 0.38 |
| Experiment | —[a] | 0.33 (0.04)[b] |

The PZ-SIC and M06-HF energies were obtained from Ref. 1, and CCSD_CCSD(T)-SP and MP2_CCSD(T)-SP denote single-point energy calculations that were carried out with the CCSD(T) method using geometry-optimized structures obtained with CCSD and MP2, respectively.

[a]No value is shown since the experimental barrier height was not provided by Ref. 1.
[b]The experimental error in the relative energy difference between DMP-L$^+$ and DPM-D$^+$ is 0.04 eV.

Before proceeding to a final discussion on the PZ-SIC transition-state geometry, we briefly discuss the accuracy of our CCSD(T) calculations, which we used as high-level calculations to benchmark both the PZ-SIC and DFT methods discussed above. First, to check for possible non-dynamical correlation effects in our CCSD(T) calculations, we computed the T1 diagnostic[3,14] for the DMP-D$^+$, DMP-L$^+$, and transition-state structures, which resulted in T1 values less than 0.031 (T1 values greater than 0.044 for open-shell systems indicate that a multi-reference electron correlation method is necessary[14]). Next, to address any possible basis-set convergence issues, we also carried out large-scale CCSD(T)-F12/cc-pVTZ (coupled-cluster with single and double excitations and perturbative triples with explicitly-correlated F12 corrections)[15] calculations – these methods exhibit dramatic improvements in basis-set convergence since they are constructed from a wavefunction that depends explicitly on the interelectronic coordinates (i.e., results of quintuple-zeta quality have been obtained with CCSD(T)-F12 methods, even when triple-zeta basis sets were used[15,16]). Nevertheless, our explicitly correlated CCSD(T)-F12 calculations are in full agreement with our CCSD(T) calculations by producing an extremely small energy barrier (~

0 eV) and a relative energy difference of 0.41 eV between the DMP-D$^+$ and DMP-L$^+$ structures. Finally, the CCSD(T) method is often referred to as the gold standard of quantum chemistry (whereas the performance of the PZ-SIC functional is much less known), and both the CCSD(T) and CCSD(T)-F12 barrier heights are extremely small (essentially barrierless and in agreement with DFT), in stark contrast to the significantly larger 0.2 eV barrier obtained from the PZ-SIC approach.

Finally, we discuss a few discrepancies regarding the transition-state geometries obtained from the PZ-SIC versus the wavefunction-based approaches. We obtained our CCSD- and MP2-optimized transition-state geometries using the Synchronous Transit-Guided Quasi-Newton (STQN)[15] method which uses a linear/quadratic synchronous transit approach to converge towards a transition-state geometry. Upon convergence, we obtained CCSD- and MP2-optimized transition-states (both exhibiting $C_s$ point-group symmetries) that closely resembled each other. However, due to the extreme computational cost of an open-shell CCSD vibrational frequency analysis, we carried out a vibrational frequency analysis on the MP2-optimized geometry (which, again, closely resembled the CCSD-optimized transition-state geometry) and obtained a single imaginary harmonic frequency of 301.18$i$ cm$^{-1}$ that connected the DMP-D$^+$ and DMP-L$^+$ structures along the potential energy curve depicted in Fig. 1 (Weber and co-workers did not carry out a vibrational frequency analysis in their study). Both the CCSD- and MP2-optimized transition-state Cartesian coordinates are provided in the Supplementary Tables 2 - 7. In contrast to the $C_s$ point-group symmetries of the transition-states obtained from MP2/CCSD, the PZ-SIC transition state is somewhat distorted and possesses a lower $C_1$ symmetry. The carbon-nitrogen bond lengths in the PZ-SIC transition-state structure are 0.1 Å smaller than those in the CCSD-optimized transition state; however, the most significant difference between the PZ-SIC and CCSD geometries were the dihedral angles of the methyl hydrogens relative to the DMP$^+$ molecule ring, which differed by as much as 33° between the two methods.

In conclusion, we have carried out new high-level CCSD(T) analyses on the DMP$^+$ cation to investigate the surprising claim that "all DFT functionals commonly used today, including hybrid functionals with exact exchange, fail to predict a stable charge-localized state" for this relatively simple system. Our new high-level CCSD(T) analyses on the DMP$^+$ cation show that DFT actually performs quite well for this system, whereas the PZ-SIC approach used by Weber et al. is the outlier that is inconsistent (and predicts a significantly larger barrier height), compared to the highly-accurate CCSD(T) benchmarks. Although the experiments by Weber et al. appear to give excellent agreement with their PZ-SIC approach, it should also be noted that their rationale for a charge-localized state of the DMP-L$^+$ cation was inferred from time-resolved measurements of the Rydberg states of the cation rather than the ground-state potential energy surface of the cation itself. While there are certainly cases where self-interaction corrections are essential for obtaining correct results in pathological chemical systems, the potential energy surface of the diamine cation, unfortunately, is not one of them.

**References**

1   Cheng, X., Zhang, Y., Jónsson, E., Jónsson, H. & Weber, P. M. Charge localization in a diamine cation provides a test of energy functionals and self-interaction correction. *Nat. Commun.* **7**, 11013, (2016).
2   Perdew, J. P. & Zunger, A. Self-interaction correction to density-functional approximations for many-electron systems. *Phys. Rev. B* **23**, 5048-5079, (1981).
3   Gudmundsdóttir, H., Zhang, Y., Weber, P. M. & Jónsson, H. Self-interaction corrected density functional calculations of molecular Rydberg states. *J. Chem. Phys.* **139**, 194102, (2013).
4   Zhang, Y., Weber, P. M. & Jónsson, H. Self-interaction corrected functional calculations of a dipole-bound molecular anion. *J. Phys. Chem. Lett.* **7**, 2068-2073, (2016).
5   Lehtola, S., Head-Gordon, M. & Jónsson, H. Complex orbitals, multiple local minima, and symmetry breaking in Perdew–Zunger self-interaction corrected density functional theory calculations. *J. Chem. Theory Comput.* **12**, 3195-3207, (2016).

6	Lehtola, S., Jónsson, E. Ö. & Jónsson, H. effect of complex-valued optimal orbitals on atomization energies with the Perdew–Zunger self-interaction correction to density functional theory. *J. Chem. Theory Comput.* **12**, 4296-4302, (2016).
7	Seo, J. *et al.* The impact of environment and resonance effects on the site of protonation of aminobenzoic acid derivatives. *Phys. Chem. Chem. Phys.* **18**, 25474-25482, (2016).
8	Melander, M., Jónsson, E. Ö., Mortensen, J. J., Vegge, T. & García Lastra, J. M. Implementation of constrained DFT for computing charge transfer rates within the projector augmented wave method. *J. Chem. Theory Comput.* **12**, 5367-5378, (2016).
9	Skúlason, E. & Jónsson, H. Atomic scale simulations of heterogeneous electrocatalysis: recent advances. *Adv. Phys. X* **2**, 481-495, (2017).
10	Zhang, Y., Deb, S., Jónsson, H. & Weber, P. M. Observation of structural wavepacket motion: the umbrella mode in Rydberg-excited N-methyl morpholine. *J. Phys. Chem. Lett.* **8**, 3740-3744, (2017).
11	Ischenko, A. A., Weber, P. M. & Miller, R. J. D. Capturing chemistry in action with electrons: realization of atomically resolved reaction dynamics. *Chem. Rev.* **117**, 11066-11124, (2017).
12	Zhang, Y., Jonsson, H. & Weber, P. M. Coherence in nonradiative transitions: internal conversion in Rydberg-excited N-methyl and N-ethyl morpholine. *Phys. Chem. Chem. Phys.* **19**, 26403-26411, (2017).
13	Kao, D.-y. *et al.* Self-consistent self-interaction corrected density functional theory calculations for atoms using Fermi-Löwdin orbitals: optimized fermi-orbital descriptors for Li–Kr. *J. Chem. Phys.* **147**, 164107, (2017).
14	Rienstra-Kiracofe, J. C., Tschumper, G. S., Schaefer, H. F., Nandi, S. & Ellison, G. B. Atomic and molecular electron affinities: photoelectron experiments and theoretical computations. *Chem. Rev.* **102**, 231-282, (2002).
15	Adler, T. B., Knizia, G. & Werner, H.-J. A simple and efficient CCSD(T)-F12 approximation. *J. Chem. Phys.* **27**, 221106, (2007).
16	Oviedo, M. B., Ilawe, N. V. & Wong, B. M. Polarizabilities of π-conjugated chains revisited: improved results from broken-symmetry range-separated DFT and new CCSD(T) benchmarks. *J. Chem. Theory Comput.* **12**, 3593-3602, (2016).

**Data availability**

The authors declare that all other data supporting the findings of this study are available within the paper (and its supplementary information files).

**Acknowledgements**

Preliminary calculations by Niranjan V. Ilawe are gratefully acknowledged. This work was supported by the U.S. Department of Energy, Office of Science, Early Career Research Program under Award No. DE-SC0016269.

**Author contributions**

B.M.W. conceived and designed the research. Z.A.A. and F.W.A. carried out the calculations and analyzed the data. Z.A.A., F.W.A., and B.M.W. wrote the paper with feedback from all co-authors.

**Additional Information**

**Supplementary Information** accompanies this paper at http://www.nature.com/naturecommunications

**Competing interests:** The authors declare no competing interests